\documentclass[]{elsart}
\usepackage[dvips]{graphicx}
\begin{document}

\begin{frontmatter}

\title{Bose-Glass behaviour in 
Bi$_{2}$Sr$_{2}$Ca$_{1-x}$Y$_{x}$Cu$_{2}$O$_{8}$ crystals 
with columnar defects: \\
experimental evidence for \\
variable-range hopping}
\author[Tours]{J.C. Soret},
\author[Tours]{L. Ammor},
\author[Tours]{V. Ta Phuoc},
\author[Tours]{R. De Sousa},
\author[Tours]{A. Ruyter},
\author[Tours]{A. Wahl},
\author[Crismat]{G. Villard}
\address[Tours]{Laboratoire d'Electrodynamique des Mat\'{e}riaux
Avanc\'{e}s \\ 
Universit\'{e} F. Rabelais - UFR Sciences - Parc de
Grandmont - 37200 Tours France}
\address[Crismat]{Laboratoire CRISMAT - UMR 6508 associ\'{e} au CNRS, \\
ISMRA et Universit\'{e} de Caen, \\
Boulevard du Mar\'{e}chal Juin - 14050 Caen Cedex France}
\begin{abstract}
We report on vortex transport in
Bi$_{2}$Sr$_{2}$Ca$_{1-x}$Y$_{x}$Cu$_{2}$O$_{8}$ crystals irradiated at
different doses of heavy ions. We show evidence of a flux-creep
resistivity typical of a variable-range vortex hopping mechanism as
predicted by Nelson and Vinokur.
\end{abstract}
\end{frontmatter}
\section{Introduction}
Disorder plays a crucial role in the static and dynamic properties of
the mixed state of high-Tc cuprates \cite{kr:1}. A main feature of the
diagram phase is the phase transition separating a high temperature
vortex liquid-like state from a low temperature vortex solid-like state.
In particular, the nature of the latter depends on the strength and the
dimensionality of disorder. In the presence of an unidimensional
disorder, Nelson and Vinokur predict a Bose-glass \cite{kr:2}. In this
paper, we shall concern ourselves with the variable-range vortex hopping
transport predicted in a Bose-glass.

\section{Experimental method}
We studied two Bi$_{2}$Sr$_{2}$Ca$_{1-x}$Y$_{x}$Cu$_{2}$O$_{8}$ crystals
with $x=0$ (sample A) and $x=0.36$ (sample B) whose the zero-field
critical temperature T$_{c}$ are 89.1K and 91K respectively. Both
samples approximately had dimensions 1$\times$1$\times$0.030 mm$^{3}$
and were together irradiated with 5.8 GeV Pb ions at the {\it Grand
Acc\'{e}l\'{e}rateur National d'Ions Lourds (GANIL)} at Caen, France.
Irradiation doses expressed in terms of a matching field B$_{\phi}$ were
1.5T (sample A) and 0.75T (sample B). In both cases the incident beam
was parallel to the c-axis of the crystal. The resulting damage consists
of parallel columnar defects of radius $c_{0}\approx 40$\AA\  throughout
the thickness of the sample.
\newline
Isothermal current-voltage (I-V) curves were recorded by using a dc four
probe method with a sensitivity better than 10$^{-10}$V and a
temperature stability better than 5mK. The magnetic field was aligned
parallel to the columnar defects.

\section{Results and discussion}
We have examined in some detail the behaviour of I-V curves in 
magnetic fields less than B$_{\phi}$. A detailed discussion of data in 
the framework of the Bose-glass melting theory \cite{kr:2} will be 
published in a separate paper.
\newline
At low temperature, we find in the limit of weak currents a non-linear
thermally actived flux-creep resistivity
\begin{equation}
\rho=\rho_{0}\exp(-\frac{c}{T}(J_{0}/J)^{\mu})  \label{eq:1}
\end{equation}
where the glassy exponent $\mu$ is independent of B and takes the value
$\sim1/3$. We have determined $\mu$ as was shown in a former paper
\cite{kr:3}.  Such barrier energies at low current is indeed the very
signature of variable-range vortex hopping\cite{kr:2}.
\newline
To test the accuracy of the above view, we estimate from the theory
\cite{kr:1,kr:2} the energy scale and the characteristic current in Eq.
(\ref{eq:1}). The energy scale $c$ is given by
\begin{equation}
	c=2d\sqrt{\tilde{\epsilon_{1}}U}  \label{eq:2}
\end{equation}
with $d \approx \sqrt{\phi_{0}/B_{\phi}}$ the mean distance between
columns, $\tilde{\epsilon_{1}} \approx
\epsilon\epsilon_{0}\ln(a_{0}/\xi_{ab})$ the tilt modulus where
$\epsilon$ is the anisotropic parameter,
$\epsilon_{0}=\phi_{0}^{2}/(4\pi\mu_{0}\lambda_{ab}^{2})$ is the line
tension and $a_{0} \approx \sqrt{\phi_{0}/B}$ is the vortice-lattice
constant, and $U=U_{0}f(T/T^{\ast})$ where $U_{0}$ is the mean pinning
energy,
$T^{\ast}=\max(c_{0},\sqrt{2}\xi_{ab})\sqrt{\tilde{\epsilon_{1}}U_{0}}$
is the energy scale for the pinning and $f(x)=x^{2}/2\exp(-2x^{2})$
accounts for thermal effects. In Eq. (\ref{eq:1})
$J_{0}=1/(\phi_{0}g(\tilde{\mu})d^{3})$ where $g(\tilde{\mu})$ denotes
the density of pinning energies at the chemical potential. Note that
$g(\epsilon)$ is normalized such that $\int_{-\infty}^{+\infty}
g(\epsilon)\,d\epsilon=1/d^{2}$. Although a form of $g(\epsilon)$ is not
yet available, an estimate of $g(\tilde{\mu})$ can be done in terms of
$\gamma$ the energy dispersion of pinning energies. Such an intrinsic
dispersion originates from the intervortex repulsion and the presence of
disorder. 
The field
independent value of the glass exponent $\mu \approx 1/3$ suggests that
$\gamma$ is dominated by the disorder effect, and thus 
we can consider $g(\epsilon)$ as a smooth function near
$\tilde{\mu}$ \cite{kr:4}. Then, $g(\tilde{\mu})=1/(d^{2}\gamma)$, wherefrom we
obtain
\begin{equation}
	J_{0} \approx \gamma/(\phi_{0}d).  \label{eq:3}
\end{equation}
In this case, one expects $\gamma$ to be given by the combination of the energy
dispersion arising from the structural disorder and the one, $\gamma_{i}$,
which results from some on-site disorder \cite{kr:1},
\begin{equation}
       \gamma \approx \gamma_{i} + t_{d}.   \label{eq:4}
\end{equation}
An estimate for the energy dispersion due to the structural disorder, i.e., the
randomness in the distance between the irradiation tracks, has been made in
terms of the hopping matrix element connecting sites separated by the typical
distance d \cite{kr:2},
\begin{equation}
       t_{d} \approx
       2\sqrt{2/\pi}\frac{U}{\sqrt{c^{\prime}/T}}\exp(-\frac{c^{\prime}}{T})
       \label{eq:5}
\end{equation}
where $c^{\prime} = c/\sqrt{2}$. On the other hand,
assuming the variations in the defect diameters induced by the 
dispersion of the ion beam as the first cause for random on-site energies, 
we approximate $\gamma_{i}$ to the width 
of the distribution of pinning energies
\begin{equation}
	\tilde{P}(U_{k})=P(c_{k})\frac{dc_{k}}{dU_{k}}  \label{eq:6}
\end{equation}
where the pinning energies $U_{k}$ and the defect radius $c_{k}$ are 
related to one another through the formula \cite{kr:2}
\begin{equation}	
	U_{k}=\frac{\epsilon_{0}}{2}\ln\left[1+\left(c_{k}/
	\sqrt{2}\xi_{ab}\right)^{2}\right]  \label{eq:7}
\end{equation}
and $P(c_{k})$ define a distribution of defects, such that
$N(c_{k})=\int_{0}^{c_{k}} P(c_{k}^{\prime}) dc_{k}^{\prime}$ is the
number of columns per unit area with radius less than $c_{k}$ and
$P(c_{k})$ is normalized such that $N(+\infty)=1/d^{2}$.
\newline 
\begin{figure}[h]
\begin{center}
\includegraphics[scale=.8]{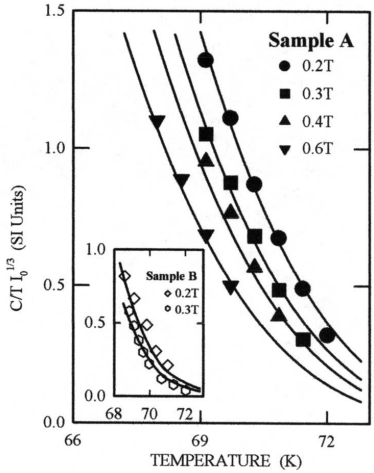}
\caption{$\frac{c}{T}I_{0}^{1/3}$ vs $T$}
\label{fg:1}
\end{center}
\end{figure}
\newline
The solid lines shown in Fig. \ref{fg:1} are a fit of
$\frac{c}{T}J_{0}^{1/3}$ to data. The exponent 1/3 accounts for the
value of the glassy exponent $\mu$. We have used the temperature
dependence as predicted in the two-fluid model for the in-plane
penetration depth i.e.,
$\lambda_{ab}(T)=\lambda_{0}\left[1-(T/T_{c})^{4}\right]^{-1/2}$, and we
have chosen a Gaussian distribution for $P(c_{k})$ centered at $c_{0}$,
with a width $w$. Indeed, Fig. \ref{fg:1} shows that the theoretical
curves and experimental data quantitatively agree. It should be noted
that the model fits above $B_{\phi}/2$ fail. Such a fit results in
realistic values for fitting parameters as shown on the table \ref{tb:1}
where $\xi_{0}$ denotes the in-plane coherence length at $T=0$.
\newline
\begin{table}[h]
\caption{Fitting parameters}
    \begin{tabular}{c|c c c c c}
        \hline
        & $c_{0}$(\AA) & $w/c_{0}$(\%) & $1/\epsilon$ & 
        $\lambda_{0}$(\AA) & $\xi_{0}$(\AA)  \\
        \hline
        Sample A & $45\pm 2$ & $17.5\pm 2.5$ & $50\pm 2$ & $1790\pm 
        20$ & $17\pm 2$  \\
        Sample B & $45\pm 2$ & $16\pm 2.5$ & $52\pm 2$ & $1910\pm 
        20$ & $15\pm 2$  \\
        \hline
     \end{tabular}
\label{tb:1}
\end{table}
\newline
Table \ref{tb:1} shows two significant points. First, the mean value of
defect diameters $2c_{0}\approx 90$ \AA\, with a dispersion of
$w/c_{0}\approx 17.5$ \% is typical of damage track diameters produced
in Bi-based compounds with 5.8 Gev Pb ions \cite{kr:5}.
Second, the only superconducting parameter
variable is the penetration depth $\lambda_{0}$. If one assumes that
$\lambda_{0}$ is independent of $B_{\phi}$, it clearly appears that
$\lambda_{0}$ increases with $x$ as expected for an underdoped sample
\cite{kr:6,kr:7}. Finally, we make a comparison between $\gamma_{i}$ and 
$t_{d}$. We obtain temperature dependent ratios of $\gamma_{i}$ to $t_{d}$ of
order 100. According to the Eq. \ref{eq:4}, this means that the energy
dispersion mainly arises from the on-site disorder. 

\section{Conclusion}
We find striking evidence (for $B < B_{\phi}/2$) for a variable-range
vortex hopping mechanism originating in the random pinning energies
induced by the energy dispersion of the ion beam.

\begin{ack}
We would like to thank H. Choplin for technical support.
\end{ack}


\begin{thebibliography}{00}
\bibitem{kr:1}  G. Blatter, M. V. Feigel'man, V. B. Geshkenbein, A.I. 
Larkin and V. M. Vinokur, {\em Rev. Mod. Phys.} {\bf 66}, 
1125 (1994).

\bibitem{kr:2}  D. R. Nelson and V. M. Vinokur, {\em Phys. Rev.} 
{\bf B48}, 13 060 (1993). 

\bibitem{kr:3}  V. Ta Phuoc, A. Ruyter, L. Ammor, A. Wahl, J. C. Soret 
and Ch. Simon, {\em Phys. Rev.} {\bf B56}, 122 (1997).

\bibitem{kr:4}  U. C. T\"{a}uber and D. R. Nelson, {\em Phys. 
Rev.} {\bf B52}, 16 106 (1995).

\bibitem{kr:5}  S. Hebert, V. Hardy, M. Hervieu, G. Villard, Ch. Simon
and J. Provost, to be published in {\em Nucl. Instr. and Meth.} {\bf B}.

\bibitem{kr:6}  G. Villard, D. Pelloquin, A. Maignan and A. Wahl, 
{\em Physica} {\bf C278}, 11 (1997).

\bibitem{kr:7}  G. Villard, D. Pelloquin and A. Maignan,
to be published in {\em Phys. Rev.} {\bf B58}.
\end{thebibliography}
\end{document}